# Boiling, steaming or rinsing?

## (physics of the Chinese cuisine)

*Andrey Varlamov[1], Zheng Zhou[2], Yan Chen[2]*

[1]CNR-SPIN, Viale del Politecnico 1, I-00133 Rome, Italy; [2]Department of Physics,

Fudan University, Shanghai, China

*Abstract: Some physical aspects of Chinese cuisine are discussed. We start from the cultural and historical particularities of the Chinese cuisine and technologies of food production. What is the difference between raw and boiled meat? What is the difference in the physical processes of heat transfer during steaming of dumplings and their cooking in boiling water? Why is it possible to cook meat stripes in a "hot pot" in ten seconds, while baking a turkey requires several hours? This article is devoted to discussion of these questions.*

Globalization, rapidly taking place in the world, is vividly manifested by the ubiquitous availability of dishes of various cuisines from all over the world. Of course, as a rule, this is just some semblance of true masterpieces of culinary art: in addition to the skill of the cook, the creation of the latter requires the corresponding products. As our familiar Italian gastronomic critic, Sergio Grasso, says, "food does not go to a person, this person should travel to food."

Chinese cuisine is one of the richest and most interesting cuisines in the world. Here everything, or, well, almost everything, can be eaten. And the ways of cooking are very different. One of the authors (AV) had a chance during his visit to Shanghai to open a small door to this wide world and, under the guidance of the other two authors, make first steps in it, being interested not only in the exotic tastes but also in underlined unusual physical processes.

Speaking about the Chinese cuisine, the first that comes in mind, probably, are the dumplings, the most worldwide popular Chinese dish, which you can eat in Chicago, Canberra and Moscow. Dumplings are honored all over China, especially in Jiangnan (close to Yangtzi Delta) region of China. Three types of dumplings are commonly seen, especially in Shanghai and Suzhou, namely, **Siaulon Pau**, **Santsie Moedeu** and **Wonton**. Made of a thin dough, filled with pork, fat of which melts into soup when cooked, and served with Chinkiang vinegar, they are pretty much the same. The most significant difference is the method of their heat treatment. **Santsie Moedeu** (large size dumpings) are pan-fried, this process is called in Shanghai **"Santsie"**. **Wonton** dumplings are boiled in 100 $^{0}$C, whereas **Siaulon** dumplings are processed also at 100°C, but in the atmosphere of saturated steam in small bamboo steaming baskets which are called **"Siaulon"**.

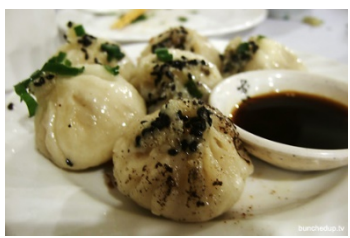
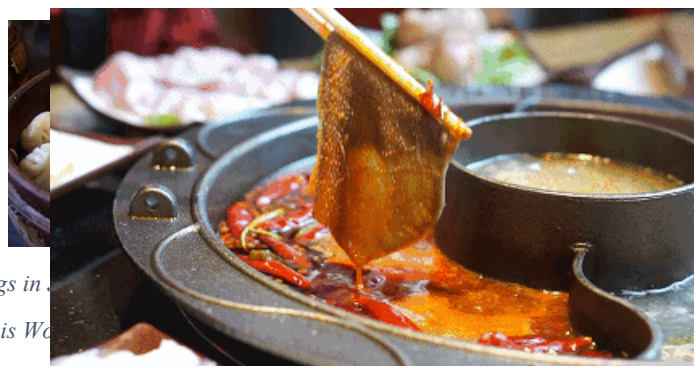

*Figure 1. The three kinds of dumplings in ... one is Siaulon Pau and the right one is W...*

*Figure 2. Cooking in the hotpot*

**Hotpot** is also a popular type of Chinese cuisine. Originated in Mongolia more than 1,000 years ago and gaining its popularity in times of the Qing Dynasty[1] all over the country, hotpot boasts a profound history. During its spread, hotpot has been diversified into many variations. **Beijing hotpot** lays particular emphasis on the soup base and sauces, **Chongqing hotpot** boasts a stimulating and refreshing "*Ma La*" ("麻辣", "numb and spicy") flavor, and **Chaoshan hotpot** is famous for its deliberately-prepared thin-cut mutton, named "*Shuan Yangrou*" ("涮羊肉") in Chinese.

When enjoying hotpot, one put ingredients such as beef balls, fish balls, crab meat, or vegetable slices into the elaborately prepared soup base and wait for it to be done. After picking it up and dipping it in the sauce, delicious food is ready to eat. The whole process is called "*Zhu*" ("煮", "to boil") in Chinese, and takes 5-10 minutes or so, being applied to meat balls and vegetables. Remarkably, that in the time between, another process can be used to get different kind of food in the same hotpot, but much faster. It is called in Chinese **"Shuan"** ("涮", "rinse" or "instant-boil"). It consists of soaking the thin-cut sliced beef or mutton in the boiling soup. Surprisingly that only in 10 seconds the sliced beef changes it color from pink to white or gray, indicating the slice is ready to eat. The beef slice becomes ready even without being go out between the chopsticks.

---

[1] The Qing dynasty was the last imperial dynasty of China, established in 1636 and ruling China from 1644 to 1912 with a brief, abortive restoration in 1917. It was preceded by the Ming dynasty and succeeded by the Republic of China.

Today, cooking has become not only a giant industry, not just art, but also a vast field of science. Here, biology, chemistry, physics, economics, ethics and much more intersect. The tasks of this science are infinite. All the time, the new methods of cooking appear. We do not even try to list them - neither about the frying of meat, nor about baking turkey, nor about the preparation of a BBQ on charcoal will not be discussed here. Let's talk about the physical processes underlying cooking on the example of the dishes described above - about the physics of boiling, steaming, and "rinsing in hotpot".

## Boiling

What is the essence of the process of meat boiling? In the everyday language, raw meat should become cooked. And what does this mean "scientifically"?

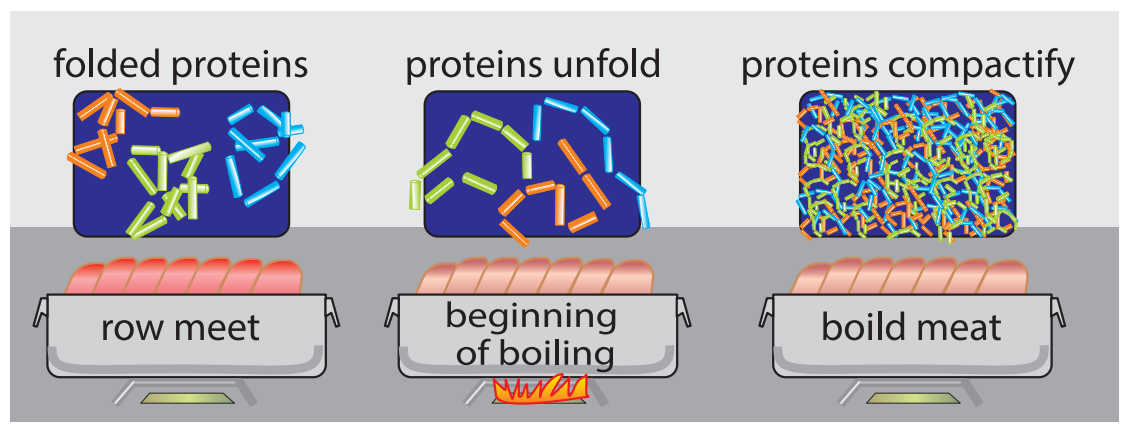

*Figure 3. Protein denaturation.*

Meat, basically, consists of complex organic macromolecules called proteins (the type of protein varies from the type of meat). In raw meat, protein molecules are in a state of entangled long chains (see Figure 3). In the course of heat treatment, the temperature rises and these chains straighten, and when the temperature reaches the specific for each type of meat value $T_d$, they are compactified into some kind of "carpet". This process is called protein denaturation. It occurs at relatively low temperatures: for meat it is 55~80°C, for fish the temperature is even lower. In any case, anyone who has ever eaten a chicken soup can be sure that boiling at 100°C turns out to be sufficient for complete compacting of proteins in the meat.

From the point of view of physics, the states of proteins in raw meat and boiled meat differ in their energy.

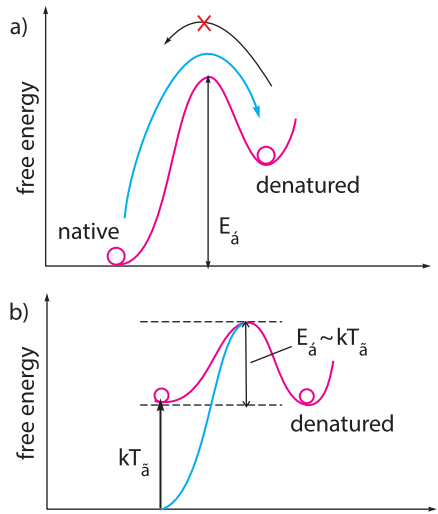

To turn the protein from its native state to the denaturated one, an energy barrier must be overcome (see Figure 4). At room temperature, this barrier is high. In the process of cooking, temperature raises. Correspondingly changes the energy of the protein, as is shown in Fig. 5. Having reached the top of the "hill", the protein falls down to the new state – a denaturated protein – the meat is cooked! This is what happens in a pot of boiling soup.[2]

So, the first task of the cook when boiling meat in terms of physics is to increase the temperature throughout the volume of the piece to at least the temperature of denaturation.

In the light of the above, we formulate the simplest model of the process of meat cooking. Let a spherically symmetric homogeneous piece of meat (radius $R$) with an initial temperature $T_0$ and a thermal conductivity coefficient $\kappa$ be placed in an environment with a fixed temperature $T_e$ is maintained. How much time does it take for the temperature of the meat in the center of the ball to reach $T_g$?

In mathematical physics, the process of heat transfer inside a sphere is described by a complex differential equation [1]

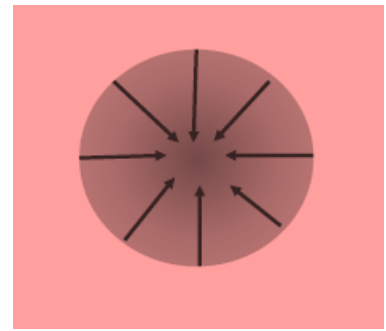

*Figure 4. Penetration of heat in the spherically symmetrical piece of meat*

---

[2] It should be noted that in recent years became fashionable to cook meat at relatively low temperatures, the so-called "sous-vide" method. The meat is placed in a thermostat with a temperature somewhat lower than that of denaturation. Each separate macromolecule lack energy alone to jump over the barrier. However, it can occasionally "borrow" it from the environment. So, gradually (it takes a long time - many hours, maybe even a day), all bulk of the meat transits into the specific denaturated state.

$$\frac{\partial T(r,t)}{\partial t} = \frac{\kappa}{\rho c} \frac{\partial}{\partial r}\left(r^2 \frac{\partial T(r,t)}{\partial r}\right) \quad (1)$$

where $T(r,t)$ is the temperature at a point $r$ at time $t$, $\kappa$ is the thermal conductivity of the meat, $\rho$ is its density, and $c$ is the specific heat. Since the water is boiling in a saucepan, the temperature at the surface of the sphere at any instant of time remains constant and equal to 100°C

$$T(r = R, \forall t) = 100°C \quad (2)$$

We took the meat from the refrigerator, so at the time when it was dropped into the water, the temperature was 4°C throughout its volume:

$$T(\forall r, t = 0) = 4^0 C \quad (3)$$

Equations (1)-(3) determine so called problem of solution of differential equation with boundary conditions. How to deal with them is well known for mathematicians and knowing the numerical values of the thermal conductivity of meat, its density and specific heat they will be able to accurately write a recipe for cooking broth. Nevertheless, let's try to figure it out by ourselves using the dimensional analysis method. The temperature of denaturation of meat coincides by the order of magnitude with the boiling point of water (differs from it by 20~25%). Therefore, we assume that the time of "delivery" of the necessary temperature to the center of the solid sphere depends only on its material parameters: the thermal conductivity of the meat, its density, specific heat and radius. Therefore, we seek the dependence of the required time on the size of the sphere in the form:

$$\tau = \kappa^\alpha \rho^\beta c^\gamma R^\delta \quad (4)$$

By comparing dimensions, we write:

$$[\tau] = [\kappa]^\alpha [\rho]^\beta [c]^\gamma [R]^\delta \quad (5)$$

The dimension of the thermal conductivity is $[\kappa] = (kg \cdot m)/(s^3 \cdot °C)$. Substituting it and dimensions of all other physical values into equation (5) and then comparing them in the right and left parts, we find: $\alpha = -1$, $\beta = \gamma = 1$, $\delta = 2$. Thus, we conclude that

$$\tau = C_0 \frac{\rho c}{\kappa} R^2 \quad (6)$$

Where $C_0$ is an unknown constant of the order of unity. Substituting the quantities $\kappa = 0.45\,W/(m \cdot °C)$, $\rho = 1.1 \times 10^3\,kg/m^3$, $c = 2.8\,kJ/(kg \cdot °C)$, we find that for the meat $\chi = \kappa/\rho c = 1.5 \times 10^{-7}\,m^2/s$. This value is called the coefficient of temperature conductivity. Consequently, a half kilogram piece of meat should be cooked for about an hour and a half. The estimate is in some way exaggerated, since we do not distinguish in its process the temperature of denaturation from the boiling point, but by the order of magnitude is correct.

Returning to the dumpling whose diameter is about 2 cm, we find that it should be cooked for several minutes, what corresponds to our life experience.

## Steaming

Now let's discuss the physical aspects of *Siaulon* dumplings preparation. Here the meat ball of the radius R (our model of the dumpling) is placed into the atmosphere of saturated steam at 100°C. The pressure here is the same as atmospheric one, i.e. is equal to 1 atm. Formally the *Siaulon* dumpling here can be considered under the same boundary conditions as the *wonton* dumpling being in the boiling water. Indeed, it is taken from the same refrigerator and is placed into environment with a temperature of 100°C. Therefore, from the point of view of a mathematician, the propagation of heat in the *Siaulon* dumpling is described by the same equation (1) with the boundary conditions (2) and (3). Therefore, if condition (2) is satisfied, then the temperature distribution inside it will be the same as for the *wonton* dumpling of the same size, and its preparation should take about the same time. However, a physicist is obliged to answer: how to ensure a temperature of 100°C on the surface of *Siaulon* dumpling?

In the case of a "wonton" this was easy: although immediately after its placement in the pan the boiling around it temporarily terminates, however, due to the high heat capacity of the water, its good enough heat conductivity, convection and constantly supplied heat to the pan, the water very quickly will boil again providing the condition (2) and hence, the required heat flow into the dumpling.

In the case of the "Siaulon", cooked in the atmosphere of the saturated steam, the mechanism of the heat transfer into the dumpling is not so evident, starting from ensure that the boundary condition (2) was satisfied. Here the heat transfer has a completely different character from that one discussed in the previous section. At the first moment, the vapor molecules near the still cold surface of the "Siaulon" locally are in the state of a strongly supersaturated vapor. They begin to condense on the surface rapidly increasing its temperature up to the temperature of the ambient, 100°C. Assuming that the temperature jump occurs in a very narrow region close to the surface, we return to the same equation (1) with boundary conditions (2) and (3). I.e., the temperature distribution inside the "siaulon" should change with time in the same way as in the case of "Wonton" boiled in the water. Consequently, the heat flux[3]

---

[3] A basic way for heat to enter the substance is thermal conduction [2,3]. When two objects with different temperature are in contact, heat flows from the one with high temperature to the one with low temperature. A basic model to study thermal conductivity is a slab of material of thickness $\Delta x$ and cross-sectional area $A$, its two ends of temperature $T_1 < T_2$. The rate of thermal conduction is measured by the heat flux density, which is the amount of heat $\Delta Q$ that flows through a unit area per unit time in the direction of temperature change

$$q = \Delta Q / A \Delta t$$

$$q(R,t) = -\kappa \left[\frac{\partial T(r,t)}{\partial r}\right]_{r=R} \qquad (7)$$

at its surface should be the same. Yet, now this flow is provided not by the thermal conductivity of the water, but by the molecules of the vapor "landing" at $1\text{cm}^2$ of the surface during 1 second:

$$q(R,t) = \Gamma\, m(t) = \Gamma\, \frac{\mu_{\text{H}_2\text{O}}}{N_A} N(t).$$

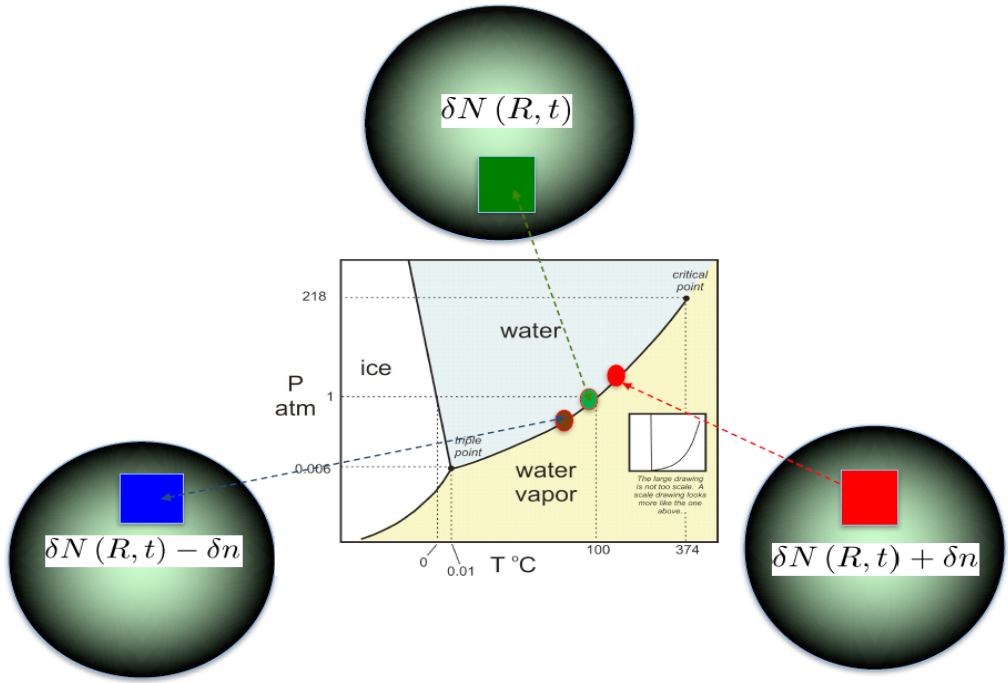

*Figure 5. A schematic representation of the process of dumpling steaming in the phase diagram of water. The environment ensures not only the heat flow but also the negative feedback suppressing its fluctuations.*

Here $\Gamma$ is the specific heat of evaporation, $N(t)$ is the number of molecules condensed per second, $m(t)$ is their mass, $N_A$ is the Avogadro number, $\mu_{\text{H}_2\text{O}}$ is the molecular mass of water. Thus, the number of molecules "landing" in the steam

---

Experimentally, $q$ is found to be proportional to be proportional to the temperature difference $\Delta T = T_2 - T_1$, and inversely proportional to the length $\Delta x$.

$$\frac{\Delta Q}{A\Delta t} = \kappa \frac{\Delta T}{\Delta x} \quad (8)$$

The coefficient $\kappa$, called thermal conductivity, again only depends on the type of substance and measures how well the substance conducts heat. More precisely, heat flux density is written in a vector $\boldsymbol{q}$ in the direction of heat flow, and $\Delta T/\Delta x$ is written in the gradient $\nabla T$ of temperature.

$$\boldsymbol{q} = -\kappa \nabla T \qquad (9)$$

The equation above is called "Fourier's law"

atmosphere at a square centimeter of "Siaulon" per one second

$$N(t) = N_A \frac{\kappa}{\Gamma \cdot \mu_{H_2O}} \left[ \frac{\partial T(r,t)}{\partial r} \right]_{r=R}$$

Perfect! Mathematicians find this number by solving a complex equation, the "Siaulon" itself "feels" what heat flux it needs to keep the temperature at 100 ° C on the surface ... It remains only to understand where from the molecules of the vapor learn how many of them should condense in a given second at a square centimeter of the "siaulon" surface.

Let's suppose that at some time $N(t) + \delta N$ instead of $N(t)$ molecules condense. The first $N(t)$ of them are hospitably absorbed by the "Siaulon" – in fact, they are necessary to keep in harmony a centigrade surface and still a cold inner part. The remaining δN are persons *"non grata"* – they were not expected here, the temperature conductivity of the "Siaulon" does not allow their heat released to penetrate into the dumpling.    What do they have to do? To take off back? Too troublesome, so they stay on the surface locally increasing its temperature (see Figure 6). As the consequence, the point that represents the vapor and water local balance moves up along their coexistence line. Let us notice that the pressure in the system remains the same, equal to 1 atm. Therefore, above the selected square centimeter, the vapor locally ceases to be saturated. As a result, the next moment here will land somewhat less number of molecules with respect to the required one. Consequently, the surface temperature will go down.

Exactly the same mechanism works also in the case of the accidental lack of condensing of the required number of molecules per second for returning of the temperature to 100°C (see Fig. 6). Such a mechanism of the self-regulation is called the negative feedback.

Let us note the important culinary difference between boiling and steaming of dumplings. In the former case water penetrates in a dumpling due to the diffusion process. Interacting with its filling it creates a tasty juice. Due to the same diffusion, but in the opposite direction, this juice partially flows out from dumpling to the ambient, transforming the water in diluted broth. In the case of dumpling steaming the ambient is the saturated steam. Condensed at the surface of a dumpling water also diffuses into its bulk, but there is no the inverse process. Hence, the juice of steamed dumpling is richer than that one in the case of boiled dumpling.

## Rinsing

At the beginning of the article we discussed the "hot pot" and the different ways of cooking meat in it. Yet, remains the mystery why the same amount of meat in the form of a meatball or thin slice differs in its cooking time tens of times.    Why oil soup base "Hongyou" in Chongqing pot boils much earlier than water soup "Qingtang" used in its Chaoshan and Beijing versions? We will try to answer these questions below.

## On cooking times

"Shuan" and "Zhu" is actually the same thermal process which was discussed in the previous sections. Here heat conductivity and convection provide a constant temperature 100°C on the surface of the object and heat enters inside it, raising the internal temperature. So, what causes this striking difference in time between rinsing the thin-cut beef and boiling the meatball? The answer requires an inspection of the thermal process of boiling.

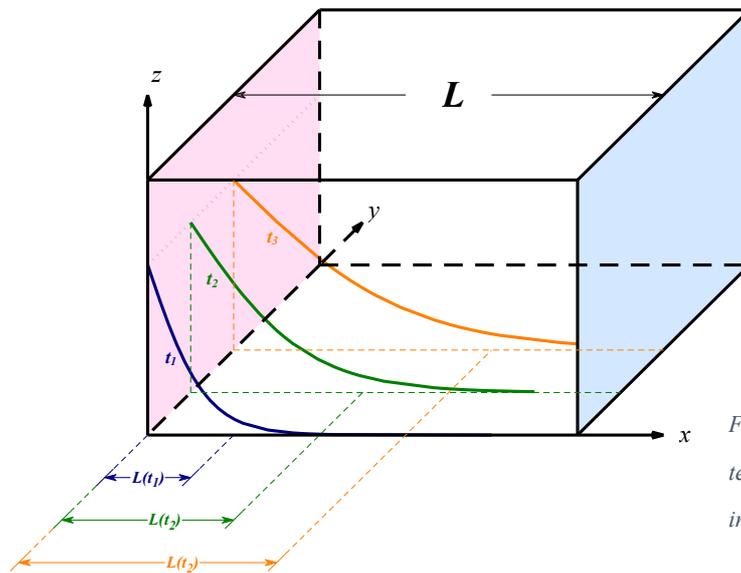

*Figure 6. A schematic of temperature penetration into a semi-space*

With the time of "Zhu" everything is clear: the way of heat propagation in the ball, which can serve as a meatball model, we already discussed (cooking of Wonton dumpling), and we can use Eq. (6) to estimate the corresponding time. The cooking time of the "Shuan" process requires separate consideration.

We consider a body occupying the semi-space $x > 0$ (see Fig. 7) We assume its density, heat capacity, and heat conductivity to be $\rho, c, \kappa$, temperature $T_0$ at the initial moment $t = 0$. Now let us change the temperature at its surface $x = 0$ to $T_1$ and fix it. The Fig. 7 shows how the temperature evolves with time in the bulk of the body. We can see that heat penetrates into the medium little by little. One can introduce characteristic length $L(t)$ to describe the propagation of the temperature front with time. Returning to Fourier law (see Eq. (9)) and replacing the thickness $\Delta x$ by the characteristic length $L(t)$, substituting $Q = mc\Delta T$, and $m = \rho AL(t)$, one finds

$$\frac{c\rho AL(t)\Delta T}{At} = \kappa\frac{\Delta T}{L(t)}$$

Solving this equation with respect to $L(t)$

$$L(t) \sim \sqrt{\frac{\kappa t}{c\rho}} = \sqrt{\chi t} \ . \quad (10)$$

Solution of the exact differential equation describing the process confirms our qualitative consideration [4] and gives an extra $\sqrt{\pi}$ coefficient in (10):

$$L(t) = \sqrt{\pi\chi t}$$

Hence, we found that heat penetrates into the medium by the square-root law of time, the parameter $\chi = \kappa/c\rho$ is called the thermal diffusivity or coefficient of temperature conductivity.[4]    The time of the establishment of the temperature $T_1$ in the slab of the thickness $L$ can then be evaluate as

$$\tau \sim L^2/\pi\chi$$

Now we can return to comparison between the rinsing time of a thin-cut beef versus the boiling time of a meatball. We consider an $a \times b$ rectangular slice of thickness $d$ and a meatball of radius $R$ out of the same volume of meat.

During the boiling process, temperature should penetrate into specimen up to farthest from the surface. Thus, the required length for thin-cut slices and meat balls are respectively half its thickness and its radius. The times for them to be done are

$$\tau_{\text{slice}} \sim \frac{(d/2)^2}{\pi\chi} \quad \tau_{\text{ball}} \sim \frac{R^2}{\pi\chi} \quad (11)$$

The ratio between their cooking times is

$$\frac{\tau_{\text{ball}}}{\tau_{\text{slice}}} \sim \left(\frac{2R}{d}\right)^2$$

The size of a thin-cut sliced beef in China typically is $a \times b = 3 \times 15$cm and its thickness is 1mm, which is marvelously thin. Out of the same amount of beef, according to the equality in volume

$$abd = 4\pi R^3/3 \ ,$$

we can make a beef ball whose radius is $R = 1.0$cm. The ratio $2R/d = 20$ and the estimated difference in cooking times

$$\left(\frac{\tau_{\text{ball}}}{\tau_{\text{slice}}}\right)_{\text{est.}} \sim 400 \quad (12)$$

Stop! As we have mentioned above the common times for the ball and the slice to be cooked are 5 minutes and 10 seconds, i.e. the realistic value for the ratio is

$$\left(\frac{\tau_{\text{ball}}}{\tau_{\text{slice}}}\right)_{\text{real.}} \sim 30 \quad (13)$$

---

[4]  Thermal diffusivity is a physical quantity that describes the rate of change (align) the temperature of the substance in a non-equilibrium thermal processes.

This number differs by one order of magnitude with (11), it is too much. The reason for this discrepancy is our assumption that all the cooking time is spent on "delivery" of the necessary temperature through the whole volume of the meat. However, we ignored the time required for carrying out of the denaturation process itself. In most of cases, this time, $\tau_{\text{denat.}}$, is so short that it can be neglected with respect to the "delivery" time. Yet, the "Shuan" process is so quick that the denaturation time cannot be neglected and has to be added to

$$\tau'_{\text{slice}} : \tau'_{\text{slice}} = \tau_{\text{denat}} + \frac{(d/2)^2}{\pi \chi} \quad (14)$$

where the temperature conductivity of the beef is taken as $\chi_{\text{beef}} = 1.5 \times 10^{-7} \text{m}^2\text{s}^{-1}$.

The value of the second term in (14) is only $\left(\frac{1}{2}d\right)^2 / \pi \chi \sim 0.5$ s, while one rinses it in the hotpot around 10 seconds. It is why we conclude that all this time is required for the chemical reaction of denaturation i.e. $\tau_{\text{denat}} \sim 10$s. Of course, this value is negligibly small compared with required to increase the temperature throughout the entire meatball volume above the denaturation point, but it turns out to be dominant when rinsing a slice of meat in a hot pot. The new estimation

$$\tau_{\text{slice}} \sim 10\text{s} \quad \tau_{\text{ball}} \sim 220 s$$

agrees well with Chinese experience.

*Table 1. Comparison among three ways of cooking mentioned above – boiling, steaming and rinsing*

| | Boiling | Steaming | Rinsing |
|---|---|---|---|
| **Size** | Radius $R$ | Radius $R$ | Thickness $L \ll a, b$ |
| **Cooking time** | $\tau \sim R^2/\pi\chi$ | $\tau \sim R^2/\pi\chi$ | $\tau \sim L^2/\pi\chi$ |
| **The Way of heat transfer** | Thermal conductivity & natural convection | Latent heat of the condensing steam | Thermal conductivity & forced convection |
| **Environment** | Boiling water/diluted broth[5] | Saturated steam | Boiling soup-base |

## On soup base

Besides the process of boiling sliced beef itself, the soup-base also contains some physics in it. First of all, it differs strongly from the diluted aquatic broth formed during dumplings boiling (see above). Soup-base for the hot pot is prepared in advance, mixing water with oil and other ingredients. Below we discuss some of its specific properties having also interesting physical grounds.

---

[5] During the process of boiling, a fraction of juice forming in the dumpling due to penetration of the water in it, returns in diffusive way back to the water ambient transforming it into diluted broth.

"Qingtang" soup base commonly is used in the Beijing and Chaoshan hotpots. It constitutes of the mix of water with a small amount of oil, salt and other condiments. Contrary, for cooking in Chongqing hotpot the "Hongyou" (spicy oil) soup base serves. A large fraction of "Hongyou" soup is oil stewed with chili powder, Chinese prickly ash, etc. The remaining part is a small portion of water. If one starts to heat both soups simultaneously, the oil soup-base will start boil much earlier than the water soup. It seems to be rather strange, since the boiling point of oil (if any) is much higher than that of water. Actually, what boils here is not the oil, but the small portion of water in the soup. Hence, the Hong You soup starts to boil at the boiling point of water instead of that one of oil.

There are two reasons which explains why the "Hongyou" soup needs a shorter time to boil.

The first is that the specific heat capacities of oil and water are respectively $c_{\text{oil}} = 2 \times 10^3 \text{ J/(kg} \cdot \text{°C)}$ and $c_{\text{water}} = 4.2 \times 10^3 \text{ J/(kg} \cdot \text{°C)}$, so that half less heat is needed to raise the temperature of the oil soup base from room temperature to 100°C.

The second reason concerns heat dissipation. The process of the heat dissipation is due to both heat conduction convection processes. In the latter energy is transferred by the movement of a heated substance as a result of differences in densities of lower and higher layers.

The heat flux at the surface is loosely described by Newton's law of cooling, which claims that the latter is proportional to the temperature difference:

$$q = h \Delta T.$$

The coefficient $h$ is called heat transfer coefficient and describes how violent the convection is. The typical values of the heat transfer coefficient for boiling water and oil are respectively $2.5 \times 10^3 \sim 25 \times 10^3$ and $0.05 \times 10^2 \sim 1.5 \times 10^2$, in the unit of $\text{W} \cdot \text{m}^{-2} \text{K}^{-1}$. The significant difference in $h$ may be explained by the viscosity and the poor heat conductivity of oil. In result, the heat dissipation at the interface between the oil soup and air is two orders in magnitude less significant than that in the case of water soup.    Manman Chi![6]

---